\documentclass{article}

\usepackage{microtype}
\usepackage{graphicx}
\usepackage{subfigure}
\usepackage{booktabs} 

\usepackage{hyperref}

\usepackage[accepted]{icml2023}

\usepackage{amsmath}
\usepackage{amssymb}
\usepackage{mathtools}
\usepackage{amsthm}
\usepackage{bbm}

\usepackage[capitalize,noabbrev]{cleveref}

\theoremstyle{plain}

\theoremstyle{definition}

\theoremstyle{remark}

\usepackage[textsize=tiny]{todonotes}

\icmltitlerunning{Long-Term Rhythmic Video Soundtracker}

\begin{document}

\twocolumn[
\icmltitle{Long-Term Rhythmic Video Soundtracker}



\icmlsetsymbol{equal}{*}

\begin{icmlauthorlist}
\icmlauthor{Jiashuo Yu}{pjlab}
\icmlauthor{Yaohui Wang}{pjlab}
\icmlauthor{Xinyuan Chen}{pjlab}
\icmlauthor{Xiao Sun}{pjlab}
\icmlauthor{Yu Qiao}{pjlab}
\end{icmlauthorlist}

\icmlaffiliation{pjlab}{Shanghai Artificial Intelligence Laboratory}

\icmlcorrespondingauthor{Yu Qiao}{qiaoyu@pjlab.org.cn}

\icmlkeywords{Music Generation, Multi-Modality, Diffusion Model}

\vskip 0.3in
]



\printAffiliationsAndNotice{}  

\begin{abstract}
We consider the problem of generating musical soundtracks in sync with rhythmic visual cues. Most existing works rely on pre-defined music representations, leading to the incompetence of generative flexibility and complexity. Other methods directly generating video-conditioned waveforms suffer from limited scenarios, short lengths, and unstable generation quality. To this end, we present \underline{Lo}ng-Term \underline{R}hythmic V\underline{i}deo \underline{S}oundtracker (LORIS), a novel framework to synthesize long-term conditional waveforms. Specifically, our framework consists of a latent conditional diffusion probabilistic model to perform waveform synthesis. Furthermore, a series of context-aware conditioning encoders are proposed to take temporal information into consideration for a long-term generation. Notably, we extend our model's applicability from dances to multiple sports scenarios such as floor exercise and figure skating. To perform comprehensive evaluations, we establish a benchmark for rhythmic video soundtracks including the pre-processed dataset, improved evaluation metrics, and robust generative baselines. Extensive experiments show that our model generates long-term soundtracks with state-of-the-art musical quality and rhythmic correspondence. Codes are available at \color{magenta}{\url{https://github.com/OpenGVLab/LORIS}.}
\end{abstract}

\section{Introduction}
\label{Introduction}

\begin{figure}[t]
\centering
\includegraphics[width=0.47\textwidth]{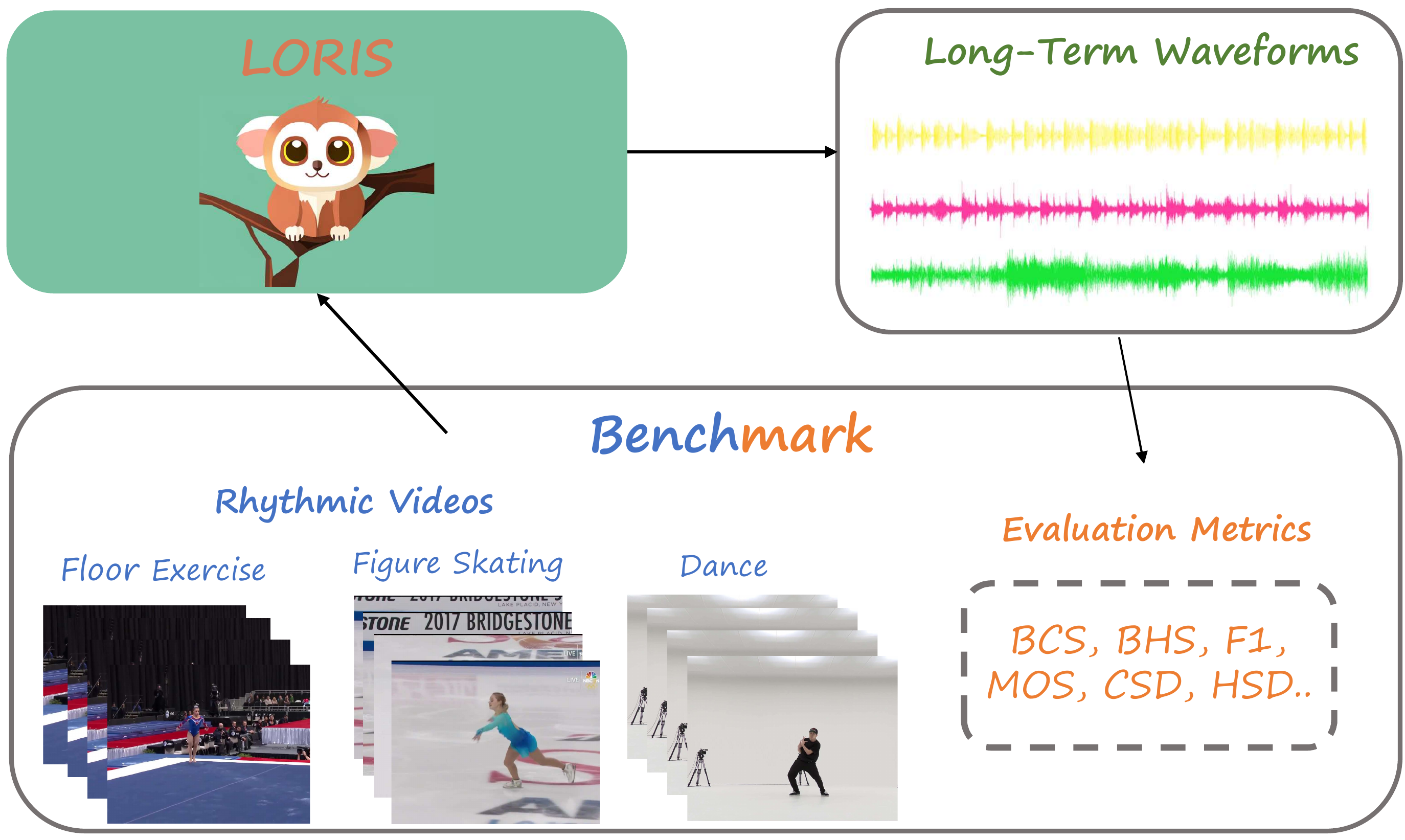}
\vspace{-3mm}
\caption{Overview of the approach. We tackle the task of generating long-term soundtracks based on given rhythmic videos. The LORIS framework is first proposed to generate long-term rhythm-correlated waveforms. We then establish a novel benchmark, including a large-scale dataset varying from dancing to sports and a set of improved metrics for long-term soundtracks.}
\vspace{-6mm}
\label{figure1}
\end{figure}

Automatic music generation has always been regarded as an iconic step towards a creative AI-generated content system. Continuous efforts~\cite{dhariwal2020jukebox, pasini2022musika, caillon2021rave, kumar2019melgan, huang2018music, von2022figaro, roberts2018hierarchical, ren2020popmag, dong2022multitrack, mittal2021symbolic} have been made to drive machines interactively generating melodious music steered by given conditionings such as genre, tempo, and style. In this paper, we work on a tightly-coupled conditioning scenario, that is, video-conditioned music generation (a.k.a. video soundtracks) which is more challenging than other conditional music generation tasks due to its cross-modality and temporal-correlated nature.  

Rhythmic video soundtracks require the model to consider the intrinsic correlations between human movements and music rhythms, and further leverage such temporal alignments as guidance for a conditional generation. To date in the literature, some works~\cite{gan2020foley, di2021video, su2020audeo, su2020multi, su2021does} investigate cross-modality soundtracks by using pre-defined symbolic musical representations such as MIDI, REMI, and piano-roll that can be autoregressively generated. However, this kind of representation is not expressive enough to cover the diverse range of sounds we hear in typical soundtracks, which hinders the model from synthesizing complex and diverse music. Recently, some advances~\cite{zhu2022quantized,zhu2022discrete} directly generate waveforms in a non-autoregressive manner, yet these works heavily rely on the computationally-expensive pre-trained music encoder~\cite{dhariwal2020jukebox}, thereby resulting in the short-length (2$\sim$6s) and low-quality consequences. Moreover, owing to the insufficiency of paired music-video data, video soundtracks are limited to the dancing scenarios, which severely restrains the model's generalizability for downstream applications.

In this paper, we introduce LORIS, the Long-Term Rhythmic Video Soundtracker to efficiently synthesize high-quality waveforms. At the heart of our model lies a latent conditional diffusion model where multiple conditionings (e.g., RGB, motions, genre) are hierarchically infused into the diffusion procedure. Specifically, we extract visual rhythms based on the cadent movement of human motions, then introduce the Hawkes Process~\cite{hawkes1971spectra, mei2017neural} on the visual rhythms to take temporal context into consideration. Besides, we also model the temporal relationship by adding a Bi-LSTM~\cite{hochreiter1997long} over the RGB embedding. These visual and motion features are conditioned via a cross-modal attention block. For the music generation part, we adopt a latent diffusion model (LDM) to encode the input waveforms into the latent feature spaces, then add and remove the Gaussian noise to/from the compressed features according to a discrete T-step schedule~\cite{karras2022elucidating}.

We also establish a comprehensive benchmark to facilitate the exploration of the rhythmic video soundtrack task. First, we build a large-scale dataset based on existing dancing and sports datasets to provide 86.43h long-term, high-quality raw videos with corresponding 2D poses, RGB features, and ameliorated audio waveforms. Next, we show the incapability of existing short-length music metrics in assessing long-term video soundtracks and propose an improved version. Finally, we conduct experiments on the established benchmark to fully evaluate LORIS on music quality and rhythmic correspondence. We show that our model, surpassing the existing methods on all metrics, can play the role of a strong baseline for the following works. In conclusion, our main contributions are three-fold:  
\vspace{-2mm}
\begin{itemize}
    \item We are the first to propose a context-aware conditional diffusion framework to perform long-term video soundtrack generation on complex rhythmic scenarios.  
    \item We propose a robust benchmark, including a large-scale rhythmic video soundtrack dataset, a set of improved evaluation metrics, and a carefully-designed baseline for the subsequent research.  
    \item Extensive experiments demonstrate that our framework is capable of generating long-term, visual-correlated musical waveforms, which benefits the creation of the musical art community.  
\end{itemize}

\section{Related Work}
\label{Related Work}

\textbf{Uni-modal Music Generation.} The family of uni-modal music generation embraces two branches. The first is in favor of using pre-defined music representations for editable music generation. Some methods~\cite{huang2018music, huang2020pop, ren2020popmag, dong2022multitrack, von2022figaro, su2020audeo} focus on transformer-based~\cite{vaswani2017attention} autoregressive models, while other advances utilize generative models such as VAE~\cite{brunner2018midi, roberts2018hierarchical}, GAN~\cite{dong2018musegan}, and DDPM~\cite{hawthorne2022multi} for the fast and conditional music synthesis. The other line of work tries to directly generate musical waveforms with less explicit constraints. WaveNet~\cite{oord2016wavenet} first shows the feasibility of autoregressively generating audio waveforms. RAVE~\cite{caillon2021rave} and Jukebox~\cite{dhariwal2020jukebox} leverage the variational autoencoder to perform high-quality audio synthesis. Some GAN-based models~\cite{kumar2019melgan, pasini2022musika} also manifest promising performance on conditional music generation.

\begin{figure*}[ht]
\centering
\includegraphics[width=0.98\textwidth]{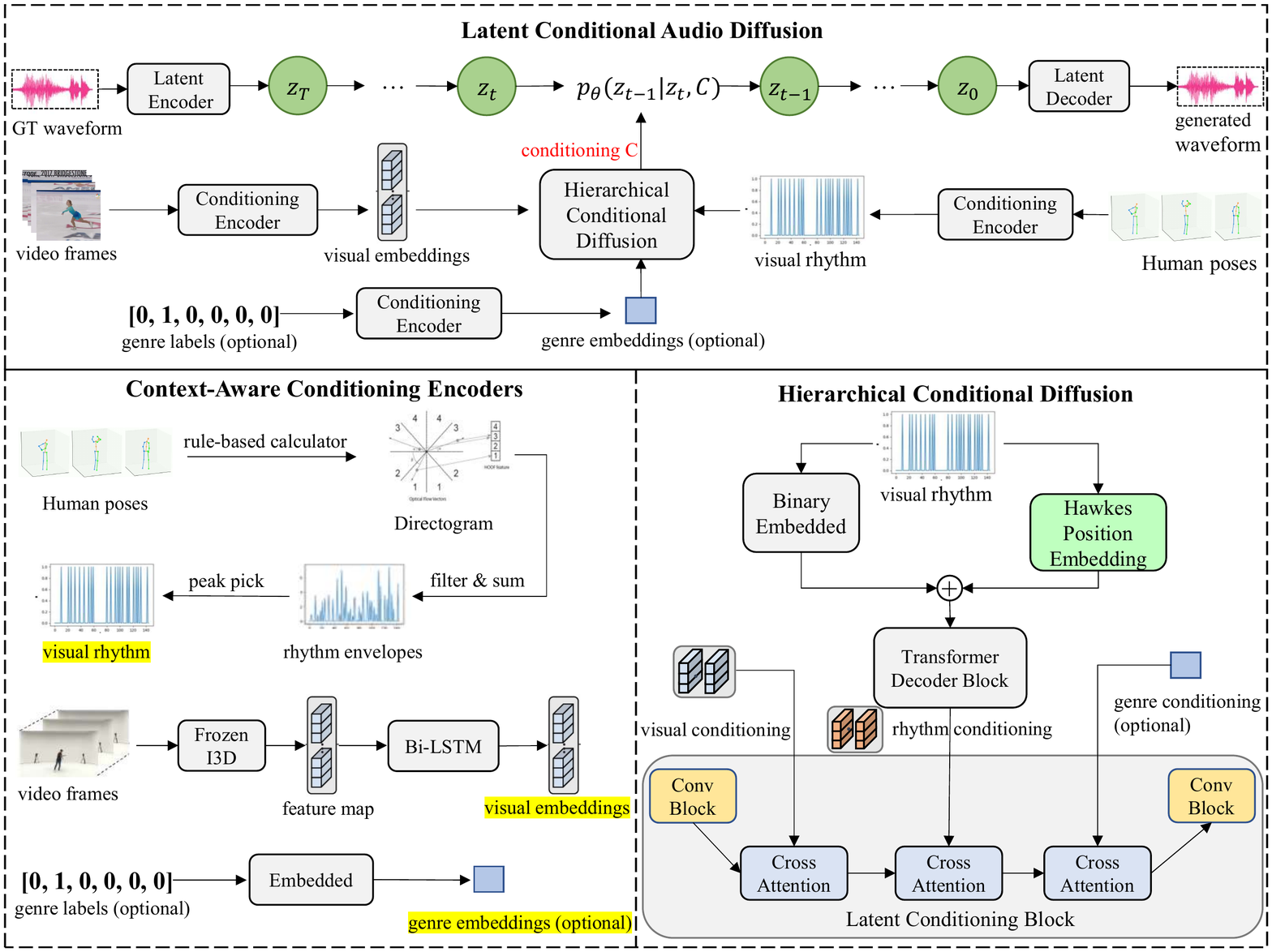}
\vspace{-2mm}
\caption{Illustration of our LORIS framework. We adopt a latent diffusion probabilistic model to perform conditional audio generation. Given an input of music-video pairs, a set of context-aware conditioning encoders first transform video frames, human poses, and categorical labels into visual embeddings, visual rhythm, and genre embeddings. Then a hierarchical conditional diffusion procedure is employed to serially attend these conditionings into the audio diffusion model, where visual rhythm is first embedded into rhythm conditioning via a Hawkes position encoding module. The entire LORIS framework is optimized jointly.}
\vspace{-2mm}
\label{figure2}
\end{figure*}

\textbf{Cross-Modal Music Generation.} To create more flexible music compositions, cross-modality generation has been studied to synthesize music correlated with inter-modality conditionings, e.g. images-to-music generation~\cite{zhang2022vis2mus, sheffer2022hear} and text-based music generation~\cite{yang2022diffsound, kreuk2022audiogen, schneider2023mo, agostinelli2023musiclm, huang2023noise2music}. These tasks usually rely on the correspondence of overall styles, while do not require fine-grained temporal alignments. More recently, several advances try to extend image-to-music generation to videos, the multi-frame scenario which needs the correlation of visual movements and musical melodies. Though some MIDI-based works~\cite{gan2020foley, su2020audeo, su2020multi} generate music in a non-regressive way, the synthesized results are highly formulated and usually mono-instrumental. More recently, D2M-GAN~\cite{zhu2022quantized} and CDCD~\cite{zhu2022discrete} directly generate video-conditioned musical waveforms. Though the results are diverse, frames are compressed into a single image as conditioning, thus temporal information is overlooked and synthesized results cannot reflect the alternation of visual movements. Besides, due to their reliance on a large music encoder~\cite{dhariwal2020jukebox}, the computation cost is extremely high, thus the music length is constrained to 2$\sim$6 seconds. Unlike prior works, our framework generates long-term waveforms (25s$\sim$50s) with affordable costs, and our context-aware design learns the audio-visual rhythmic correlation to ensure inter-modality coherence.

\textbf{Generalized Cross-Modal Generation.} Remarkable advances also exist in other inter-modality generation tasks. Text-to-image synthesis has drawn increasing attention~\cite{ramesh2022hierarchical, rombach2022high, gu2022vector, tang2022improved} in pace with the gorgeous growth of contrastive language-image pre-training~\cite{radford2021learning} and diffusion models~\cite{ho2020denoising}, where the synthesized images exhibit high-resolution quality with great diversity and compute-efficiency. Some works~\cite{singer2022make, hong2022cogvideo} also extend text-to-image to conditional video generation, while more recent pioneering methods investigate text-based pose sequences~\cite{xie2022vector} and 3D scenes~\cite{poole2022dreamfusion} generation. In this work, we refer to the latent conditional diffusion mechanism utilized by text-to-image approaches and attach more visual cues to stabilize long-term music synthesis.

\section{Methodology}
\label{Methodology}

LORIS is depicted in Figure~\ref{figure2}. Given a music-video pair, the latent diffusion model (Section~\ref{met:subsec1}) is used to synthesize auditory waveforms, and a set of conditioning encoders (Section~\ref{met:subsec2}) is designed to generate context-aware visual cues. Besides, a hierarchical conditional diffusion module (Section~\ref{met:subsec3}) is proposed to add cross-modality constraints.

\subsection{Unconditional Latent Diffusion}
\label{met:subsec1}

Inspired by the recent image-to-text generation advance~\cite{rombach2022high}, we use a similar open-source architecture audio-diffusion-pytorch~\cite{schneider2023archisound} pre-trained on a large-scale YouTube music dataset as our uni-modal diffusion backbone. Due to the numerous amount of audio sampling points, we encode the input waveforms into a compressed latent representation $z\sim p_{data}$ to lower the computation cost. Unconditional latent diffusion consists of a forward diffusion process and a reverse denoising procedure. The forward process can be regarded as a Markov chain that progressively corrupts the initial latent codes $z_0$ into Gaussian noise $z_T\sim \mathcal{N}(0, \mathbf{I})$ with a sequence of $T$ steps. In contrast, the objective of the denoising process is to reverse the Gaussian distribution to the original vectors in identity steps. The denoising error can be optimized via the L2 objective. We seek to directly predict $z$ rather than utilizing the $\epsilon$-prediction formulation~\cite{ho2020denoising}:
\begin{equation}
\small
    L_{LD} := \mathbb{E}_{z\sim p_{data}, t\sim [1, T]}\left[\lambda(\sigma_t)\|D_{\theta}(z, \sigma_t) - z\|\right]_2^2],
\end{equation}
where $D_\theta$ is the denoising network parameterized by $\theta$, $\lambda(\sigma_t)$ denotes an optional weighting function.

In practice, EDM~\cite{karras2022elucidating} is employed to improve the denoiser:
\begin{equation}
\small
D_{\theta}(z, \sigma_t) = c_{skip}(\sigma_t)z + c_{out}(\sigma_t)f_{\theta}(c_{in}(\sigma_t)z, \frac{1}{4}ln(\sigma_t)),
\end{equation}
where $c_{skip}(\sigma_t), c_{out}(\sigma_t)$, and $c_{in}(\sigma_t)$ are scaling parameters and $\lambda(\sigma_t)$ is ameliorated as $1/c_{out}(\sigma_t)^2$. Details about scaling parameters are listed in Appendix~\ref{scalingparam}.

\subsection{Context-Aware Conditioning Encoders}
\label{met:subsec2}

Previous waveform-based methods~\cite{zhu2022quantized, zhu2022discrete} share a common paradigm that compresses the temporal dimension and encodes frames into a global visual embedding $f_e\in \mathbb{R}^{1\times C}$, where C denotes the hidden dimension. Although these global features are flexible to act as conditional guidance, contextual information is overlooked, thence the model is incapable of synthesizing correlated music that responds to the change of visual contents. Such a phenomenon also elucidates why existing waveform-based methods can only tackle short-length videos. To this end, we model the temporal correspondence explicitly and construct visual conditioning $c_v$, rhythm conditioning $c_r$, and genre conditioning $c_g$ (if necessary) via different conditioning encoders.  

\noindent\textbf{Visual Conditioning.} For the visual encoder, we follow previous methods~\cite{zhu2022quantized, zhu2022discrete} that use pre-trained I3D~\cite{carreira2017quo} network as feature extractor. Differently, we do not perform feature aggregation across the temporal dimension and leverage a Bi-LSTM~\cite{hochreiter1997long} layer to capture long-range temporal dependencies:
\begin{equation}
\small
    c_v, (h, mc) = BiLSTM(Enc(i_1, i_2, ..., i_T), h_0, mc_0),
\end{equation}
where $Enc$ is the visual encoder, $I = \{i_1, i_2, ..., i_T\}$ are input visual frames, $h$ and $mc$ denote hidden state and memory cell state vectors, $h_0, mc_0$ indicate their initial state. The parameters of I3D are frozen during training while the Bi-LSTM layer is involved in optimization.  

\noindent\textbf{Rhythm Conditioning.} Several approaches have been proposed to extract dance rhythms, such as measuring the rapid changes of optical flows~\cite{davis2018visual}, performing Short-Time-Fourier Transform on human skeletons~\cite{su2021does}, or merging neural networks with traditional graphic functions~\cite{yu2022self}. Considering the commonality of rhythmic videos, we put forward an improved rule-based method to encode all frames of each video into a binary vector to represent visual rhythm points. Concretely, we first extract 2D poses $P(t, j, x, y)$ via pre-trained models, where $t$ and $j$ denote the current temporal position and key joint, $x, y$ denotes the joint coordinate, then calculate the first-order difference of as 2D motions $M(t, j, x, y)$. To comprehensively estimate the kinematic amplitude and strength, we utilize the directogram~\cite{davis2018visual}, a 2D matrix $D$ analogous to the audio spectrogram to represent the change of motions. Motions in each timestamp are first divided into $K$ bins based on their angles with x-axis by $tan^{-1}\frac{y}{x}$, and the weighted summation is computed as the directogram:
\begin{equation}
\small
    D(t, \theta) = \sum_{j}\|M(t, j)\|_2\mathbbm{1}_\theta (\angle M(t, j)),
\end{equation}
\vspace{-2mm}
\begin{equation}
\small
    \mathbbm{1}_\theta(\phi) := 
    \begin{cases}
    1, &\text{if } |\theta-\phi|\le \frac{2\pi}{K}, \\
    0, &\text{otherwise}.
    \end{cases}
\end{equation}
Similar to audio onset envelopes, we calculate the bin-wise difference of the directogram, sum all positive values in each angular column, and normalize the resulting curves into the range of $[0, 1]$ as the visual onset envelopes $O$:
\begin{equation}
\small
    O(t) = \eta(\sum_{k=1}^{K}max(0, |D(t, k)|-|D(t-1, k)|)),
\end{equation}
where $D(t, k)$ denotes the directogram volume at $t$-th time step and $k$-th bin, $\eta$ is the normalized function.

Although $O(t)$ can already be regarded as the visual rhythmic conditioning, we further employ a peak-picking strategy~\cite{bock2012evaluating} to simplify the continuous curves into discrete binary codes for the convenience of conditional generation. Specifically, the temporal point $t$ can be identified as the $i$-th visual rhythmic point only when the following prerequisites are satisfied:
\begin{equation}
\begin{aligned}
\small
    &c_r(t_i) = max(O[t_i-pre_m:t_i+post_m]),              \\
    &c_r(t_i) \geq mean(O[t_i-pre_a:t_i+post_a]) + \delta,  \\
    &t_i - t_{i-1} > \omega
\end{aligned}
\end{equation}
where $c_r(t_i)$ is the $i$-th rhythm peak in temporal position $t$, $pre_m$ and $post_m$ denote the distance of finding local maxima before and after the current position; $pre_a$ and $post_a$ indicate the distance of computing the local average, $\delta$ is the threshold that the local maxima must be above the local average. 

Finally, we get a binary vector $c_r\in \mathbb{R}^{T\times 1}$ that represents visual rhythm peaks, where 1 denotes that the current time step is one of the rhythm points. We also explain the rationale of our improved visual rhythm extraction method in Appendix~\ref{rhythmrationale}.

\noindent\textbf{Genre Conditioning.} Rhythmic videos can be categorized into different types based on their characteristics such as the choreography styles of dancing videos. Taking genre into consideration could facilitate some interesting applications like style transfer and music editing. We regard the musical genre as global conditioning and embed one-hot categorical labels $G$ into genre features via linear projection: 
\begin{equation}
\small
    c_g = Embed(G).
\end{equation}

It is noted that genre conditioning can only be utilized for datasets that include musical genre or category labels.  

\subsection{Hierarchical Conditional Diffusion}
\label{met:subsec3}
Considering the enormous success of cross-attention mechanism~\cite{vaswani2017attention} in conditional generation~\cite{ramesh2022hierarchical, rombach2022high}, we adopt such an approach to model the correlation between latent feature $z$ and conditioning $c$. One obstacle for conditional diffusion is that rhythm peaks are binary vectors, which cannot perform feature interactions due to the low dimension. Therefore, we employ a trainable linear layer $W_r$ similar to the genre encoder to project the binary vector to a rhythm embedding matrix $c_rW_r \in \mathbb{R}^{T\times C}$. We further argue that in a sequence of rhythm points, the mechanism of peak-picking determines the temporal point neighboring to rhythm peaks is unlikely to be another rhythm peak, thus we can explicitly add positional penalties to those temporal points adjacent to rhythm peaks. To this end, we introduce the Hawkes Process~\cite{hawkes1971spectra, mei2017neural, zhang2020self}, where additional temporal offsets are attached over the raw positional encoding in each dimension:
\begin{equation}
\small
    Hawkes^k(t_i) = Tri(\omega_k \times i + w_k \times t_i),
\end{equation}
where $Tri$ denotes $sin$ and $cos$ for the even and odd dimension, $\omega_k, w_k$ are the learnable parameters in the $k$-th dimension for Hawkes encoding and positional encoding, respectively, $t_i$ denotes the temporal position of the $i$-th rhythm peak. In practice, the shifted position can be computed as $i' = i + \frac{w_k}{\omega_k}t_i$, where $\frac{w}{\omega}$ for all dimensions can be regarded as a learnable parameter matrix. In this way, our model takes contextual rhythm information into consideration for more accurate rhythmic control.

Then we add the shifted positional embeddings to the rhythm embedding $\hat{c}_r$ and use a Transformer~\cite{vaswani2017attention} decoder block to perform feature integration:
\begin{equation}
\small
    \hat{c}_r = TrmDec(c_rW_r + Hawkes(t_i)),
\end{equation}
After acquiring the conditional embeddings $\{c_v, \hat{c}_r, c_g\}$, cross-modal attention is employed to interact conditional embeddings with intermediate layers of U-Net~\cite{ronneberger2015u} by computing feature similarity:
\begin{equation}
\small
    Att(c_{\alpha}, \psi_i(z)) = Softmax(\frac{W_Q^i\psi_i(z)\cdot(W_K^ic_{\alpha})^T}{\sqrt{d}}) \cdot W_V^ic_{\alpha},
\end{equation}
where $c_{\alpha}\in \{c_v, \hat{c}_r, c_g\}$ denotes the conditional embeddings, $\psi_i(z)$ denotes the $i$-th intermediate tensor of latent feature $z$, $W_Q^i, W_K^i, W_V^i$ are learnable projection matrices. 

Notably, we put conditionings serially to adapt the divergent temporal lengths of visual cues, by virtue of which we can use one cross-modal attention block to attend RGB, rhythm, and genre embeddings with far less computation cost.

Given the conditioning $C = \{c_v, \hat{c}_r, c_g\}$, the objective for the latent conditional denoising can be formulated as:  
\begin{equation}
\small
    L_{CLD} := \mathbb{E}_{z\sim p_{data}, t\sim [1, T]}\left[\lambda(\sigma_t)\|D_{\theta}(z, \sigma_t, C) - z\|\right]_2^2].
\end{equation}

\section{Benchmark}
\label{Benchmark}
\subsection{Dataset}

\begin{table*}[tb]
\centering
\small
\begin{tabular}{@{}lllll@{}}
\toprule
Dataset         & Origin                   & Scenarios        & Total length     & Segment length    \\ \midrule
AIST++$_{st}$~\cite{li2021ai}     & AIST++       & Dance            & 3.78h            & 2s, 6s            \\
Tiktok~\cite{zhu2022quantized}          & YouTube                      & Dance                                                                            & 1.55h        & 2s             \\  \midrule
Ours          & \begin{tabular}[c]{@{}l@{}}AIST++, FineGym, \\ FS1000, FisV\end{tabular} & \begin{tabular}[c]{@{}l@{}}Dance, Floor Exercise, \\ Figure Skating\end{tabular} & 86.43h       & 25s, 50s       \\ \bottomrule
\end{tabular}
\caption{Comparison between our dataset and existing video soundtrack datasets. AIST++$_{st}$ means cutting raw AIST++ videos into 2s and 6s short-length segments. Statistical results show our dataset involves more video-music pairs with more categories and longer lengths.}
\label{table:dataset}
\vspace{-3mm}
\end{table*}

 Since prior waveform-based methods tackle short-length videos, the primary obstacle for long-term soundtrack generation is the shortage of paired audio-visual data. To this end, we curate the LORIS dataset based on existing datasets, which involves 86.43h paired videos varying from dances to multiple sports events. The comparison of our dataset with existing datasets is listed in Table~\ref{table:dataset}. To be specific, our dataset incorporates three rhythmic categories: dance, figure skating, and floor exercise. The dancing videos are curated from AIST++~\cite{li2021ai} dataset, a fine-annotated subset of AIST~\cite{tsuchida2019aist}. We select all videos longer than 25 seconds and preserve their categorical labels to perform genre conditioning. Although 3D meshes and skeletons are available, we only curate the original videos. Figure skating videos are collected from FisV~\cite{xu2019learning} dataset and FS1000~\cite{xia2022skating} dataset, and floor exercise videos are from Finegym~\cite{shao2020finegym} dataset. For the sports videos, we only use the raw videos and do not utilize any annotation or provided features. 

 After curating the raw videos, we make the following pre-processes: 1). We cut off the first and last 5 seconds of each sports video, and divide these videos into 25s and 50s segments. 2). We adopt the sound source-separated framework Spleeter~\cite{spleeter2020} and employ its 2stem pre-trained model to remove vocals, commentaries, and cheers to acquire pure 16kHz musical accompanies. 3). We manually filter video splits and remove the videos with noisy audio, unseparated vocals and cheers, absent background music, and overmuch missing frames. 4). We upsample the music sample rate to 22kHz. 5). We extract the RGB features of visual frames using I3D~\cite{carreira2017quo} pre-trained on Kinetics~\cite{kay2017kinetics} and Charades~\cite{zhang2020temporal} datasets, and employ mmPose~\cite{mmpose2020} to obtain 2D skeletons using HRNet~\cite{sun2019deep} pre-trained on MS COCO~\cite{lin2014microsoft} dataset. 6). Finally, we randomly split the dataset with a 90\%/5\%/5\% proportion. To sum up, we curate 12,446 25-second paired videos, including 1,881 dancing videos, 8,585 figure skating videos, and 1,950 floor exercise videos. For the 50-second versions, our dataset includes 4,147 figure skating videos and 660 floor exercise videos.

\subsection{Evaluation Metrics}

We follow the general paradigm of previous works~\cite{zhu2022quantized, zhu2022discrete} that measure musical quality and cross-modality correspondence. For the musical quality, the subjective metrics Mean Opinion Scores (MOS) for the general quality are reported. To investigate rhythm correspondence, we use the improved versions of beats coverage scores (BCS) and beats hit scores (BHS) for evaluation. To be specific, BCS and BHS are first proposed for music-guided dance generation~\cite{davis2018visual, lee2019dancing} which measures the alignment of musical rhythms and dancing patterns. Similarly, prior dance-to-music methods~\cite{zhu2022quantized, zhu2022discrete} employ these metrics to count the aligned rhythm points of synthesized music and ground-truth music by computing the rhythm point number of generated music $B_g$, the rhythm point number of ground-truth music $B_t$, and the number of aligned rhythm points $B_a$. Then, BCS is calculated as the fraction of generated musical beats by the ground truth musical beats ($B_g/B_t$), and BHS measures the ratio of aligned beats to the ground truth beats ($B_a/B_t$). However, we found that these metrics are only suitable for short-length (2$\sim$6s) music, and two main problems emerge when evaluating long-term soundtracks: 1). the second-wise rhythm detection algorithm results in an extremely sparse vector for any long music sequence, thus the constantly low BCS and BHS values are unable to reflect the real performance. 2). BHS can easily exceed 1 if generated music involves more rhythm points than ground truth. Considering a batch involves two samples with BHS of 0.5 and 1.5, the average BHS is 1, which seems to be perfect while each sample performs unsatisfactorily. Hence, the reported value cannot reflect the real quality under such metrics. Accordingly, we make two corresponding modifications: 1. We adjust the parameters of audio onset detection algorithms~\cite{bock2012evaluating} (More details in Appendix~\ref{evamet}) to avoid sparse rhythm vectors. 2. We calculate BCS by dividing the aligned beats by the total beats from the generated music ($B_a/B_g$), by which BCS and BHS play the roles of recall and precision, respectively. Besides, we calculate the F1 scores of BCS and BHS as an integrated assessment and report the standard deviations of BCS and BHS (termed CSD and HSD, respectively) to evaluate generative stability.

\subsection{Baselines}
 
To make an exhaustive evaluation, we choose several well-performed methods with available codes as baselines. Concretely, we re-implement MIDI-based methods Foley~\cite{gan2020foley} and CMT~\cite{di2021video}, waveform-based methods D2M-GAN~\cite{zhu2022quantized} and CDCD~\cite{zhu2022discrete} on our dataset. Experimental results of the baseline methods and our LORIS framework on the established benchmark are reported in Section~\ref{Experiments}.

\section{Experiments}
\label{Experiments}

\begin{table}[htbp]
\small
\resizebox{\linewidth}{!}{
\begin{tabular}{ccccccc}
\hline
Metrics          & BCS$\uparrow$  & CSD$\downarrow$        & BHS$\uparrow$      & HSD$\downarrow$    & F1$\uparrow$     & MOS$\uparrow$      \\ \midrule
Foley            & 96.4           & 6.9                    & 41.0               & 15.0               & 57.5             & 3.2                \\
CMT              & 97.1           & 6.4                    & 46.2               & 18.6               & 62.6             & 3.6                \\
D2MGAN           & 95.6           & 9.4                    & 88.7               & 19.0               & 93.1             & 2.8                \\
CDCD             & 96.5           & 9.1                    & 89.3               & 18.1               & 92.7             & 3.1                \\ \midrule
Ours             & \textbf{98.6}  & \textbf{6.1}           & \textbf{90.8}      & \textbf{13.9}      & \textbf{94.5}    & \textbf{3.7}       \\ \hline
\end{tabular}}
\caption{Quantitative results on the LORIS$_{DA25}$ dancing subset.}
\label{table:da25}
\end{table}

\begin{table*}[htbp]
\small
\centering
\begin{tabular}{ccccccccccccc}
\hline
Subset           & \multicolumn{6}{c}{LORIS$_{FE25}$}                        & \multicolumn{6}{c}{LORIS$_{FE50}$}                      \\ \cmidrule[.08em](l{.3em}r{.3em}){2-7} \cmidrule[.08em](l{.3em}r{.3em}){8-13}
Metrics          & BCS$\uparrow$      & CSD$\downarrow$     & BHS$\uparrow$       & HSD$\downarrow$     & F1$\uparrow$    & MOS$\uparrow$    & BCS$\uparrow$      & CSD$\downarrow$     & BHS$\uparrow$       & HSD$\downarrow$     & F1$\uparrow$    & MOS$\uparrow$  \\ \midrule
Foley            & 36.0     & 36.2    & 32.3      & 30.7    & 34.1  & 3.2                & 32.6      & 38.0    & 28.4     & 32.5    & 30.4  & 3.1               \\
CMT              & 46.4     & 30.1    & 57.4      & 29.8    & 51.3  & \textbf{3.8}       & 42.3      & 32.0    & 53.8     & 31.7    & 47.4  & \textbf{3.5}      \\
D2MGAN           & 45.3     & 27.7    & 58.7      & 30.1    & 51.1  & 2.4                & 41.9      & 29.2    & 54.7     & 32.7    & 47.5  & 2.1               \\
CDCD             & 49.0     & 21.1    & 61.0      & 27.0    & 54.3  & 2.7                & 45.9      & 23.8    & 57.5     & 29.3    & 51.0  & 2.5               \\ \midrule
Ours             & \textbf{58.8}     & \textbf{19.4}    & \textbf{67.1}      & \textbf{21.1}    & \textbf{62.7}  & \textbf{3.8}       & \textbf{54.7}      & \textbf{21.6}    & \textbf{63.8}     & \textbf{24.5}    & \textbf{58.9}  & 3.4               \\ \hline
\end{tabular}
\caption{Comparison with state-of-the-art methods on the LORIS$_{FE25}$ and LORIS$_{FE50}$ floor exercise subsets.}
\label{table:fe}
\end{table*}

\begin{table*}[htbp]
\small
\centering
\begin{tabular}{ccccccccccccc}
\hline
Subset           & \multicolumn{6}{c}{LORIS$_{FS25}$}                        & \multicolumn{6}{c}{LORIS$_{FS50}$}                      \\ \cmidrule[.08em](l{.3em}r{.3em}){2-7} \cmidrule[.08em](l{.3em}r{.3em}){8-13}
Metrics          & BCS$\uparrow$      & CSD$\downarrow$     & BHS$\uparrow$       & HSD$\downarrow$     & F1$\uparrow$    & MOS$\uparrow$    & BCS$\uparrow$      & CSD$\downarrow$     & BHS$\uparrow$       & HSD$\downarrow$     & F1$\uparrow$    & MOS$\uparrow$  \\ \midrule
Foley            & 36.1     & 25.4    & 21.1               & 16.2    & 26.7  & 3.0                & 33.8     & 27.8    & 18.5           & 18.2    & 24.0  & 3.2     \\
CMT              & 37.2     & 24.9    & \textbf{73.8}      & 29.2    & 49.4  & \textbf{3.9}       & 34.6     & 26.4    & \textbf{70.7}  & 31.1    & 46.5  & \textbf{3.6}    \\
D2MGAN           & 42.8     & 26.8    & 48.9               & 25.9    & 45.7  & 2.7                & 39.9     & 28.7    & 45.7           & 27.0    & 42.6  & 2.2     \\
CDCD             & 45.6     & 27.3    & 44.6               & 22.6    & 45.1  & 2.8                & 42.1     & 29.5    & 41.6           & 24.6    & 41.8  & 2.4     \\ \midrule
Ours             & \textbf{52.2}     & \textbf{18.5}    & 57.0      & \textbf{19.8}    & \textbf{54.5}  & 3.8       & \textbf{49.3}     & \textbf{20.8}    & 54.9      & \textbf{21.4}    & \textbf{52.0}  & \textbf{3.6}     \\ \hline
\end{tabular}
\caption{Comparison with state-of-the-art methods on the LORIS$_{FS25}$ and LORIS$_{FS50}$ figure skating subsets.}
\label{table:fs}
\end{table*}

\subsection{Implementation Details}
We use audio-diffusion-pytorch-v0.0.43~\cite{schneider2023archisound} as our basic backbone. The dimension of all hidden layers is set to 1024, and the embedding size of genre labels and RGB features is also 1024. For visual rhythm extraction, the bin number $K$ is set to 10 and hyperparameters of peak picking strategy are $pre_m=3$, $pre_m=3$, $post_a=3$, and $post_a=3$. We set the threshold offset $\delta$ as 0.2 multiples by the current local maxima, and peak wait number $\omega = 1$. The audio sampling rate is set to 22050 Hz. We use AdamW~\cite{loshchilov2017decoupled} as the optimizer with $\beta_1 = 0.9$, $\beta_2 = 0.96$ and weight decay of 4.5e-2. A two-stage learning rate strategy is applied during training. Specifically, for layers in the unconditional diffusion model pre-trained by~\cite{schneider2023archisound}, we set the learning rate as 3e-6 while the initial learning rate of all other layers is 3e-3. We set a warm-up learning rate of 2e-4 for all layers in the first 1,000 training iterations. We also apply the gradient clipping with the max norm of 0.5. The entire LORIS framework is optimized jointly and we use 8 NVIDIA A100 GPUs to train our model for 100 epochs on the dancing subset, 200 epochs on the floor exercise subset, and 250 epochs on the figure skating subset. For music sampling, we employ the classifier-free guidance~\cite{ho2022classifier} to perform conditional generation with guidance scale $w = 20$. The diffusion step number during inference is set to 50 as a trade-off of music quality and inference speed.

\subsection{Main Results}

Since visual appearances and musical patterns of dance and sports datasets vary widely, we separately evaluate each rhythmic type termed as LORIS$_{DA25}$ subset for 25s dancing videos, LORIS$_{FE25}$ and LORIS$_{FE50}$ subsets for 25s and 50s floor exercise videos, and LORIS$_{FS25}$ and LORIS$_{FS50}$ subsets for 25s and 50s figure skating videos.  

Results on the 25s dancing subset are shown in Table~\ref{table:da25}, where our model outperforms all previous methods both on rhythmic coherence and musical quality. In particular, All methods show satisfactory BCS and low CSD since dancing beats are periodic and easy to perceive, thereby the rhythm points of generated music $B_g$ and GT music $B_t$ are similar. However, waveform-based methods D2M-GAN~\cite{zhu2022quantized} and CDCD~\cite{zhu2022discrete} achieve higher BHS than MIDI-based methods Foley~\cite{gan2020foley} and CMT~\cite{di2021video}, which suggests that waveforms synthesized by generative models are more flexible to perform rhythm alignment.  
Performance on sports subsets is demonstrated in Table~\ref{table:fe} and Table~\ref{table:fs}. We find that all methods perform worse in rhythmic coherence and musical quality on the sports datasets, indicating sports are more challenging rhythmic scenarios. Nevertheless, our model still achieves considerable boosts compared with CDCD~\cite{zhu2022discrete} about +8.4\%, +8.2\% F1 scores for 25s and 50s floor exercise videos, and +9.4\%, +8.2\% F1 scores for figure skating videos. These numerical results verify that our LORIS framework is capable of generating high-quality musical soundtracks with accurate rhythmic alignment for both dances and sports.  

\begin{figure*}[htbp]
\centering
\includegraphics[width=0.98\textwidth]{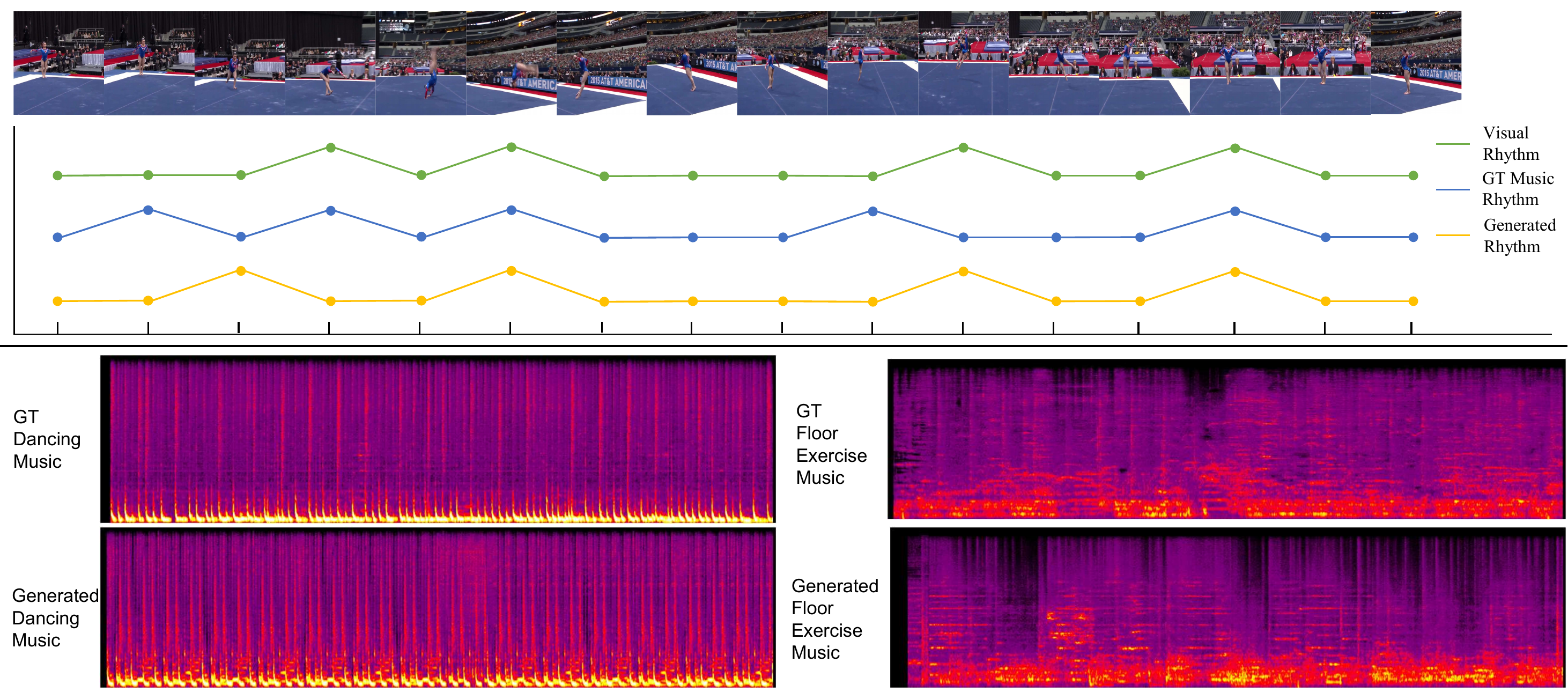}
\vspace{-2mm}
\caption{Visualizations of rhythms and musical log-melspectrograms. Examplar in the upper of the picture shows the rhythm correspondence between audio and visual rhythms, where the green curve indicates visual rhythm peaks extracted via our rule-based strategy, and blue and yellow curves denote the ground truth and generated musical rhythm points. Results show that our model synthesizes music with satisfactory rhythm coherence. The lower part is the comparison of generated and ground-truth musical log-melspectrogram, where the synthesized results lie in an analogous crest distribution with ground-truth music.}
\vspace{-4mm}
\label{figure3}
\end{figure*}

\subsection{Ablation Studies}

\begin{table}[htbp]
\small
\centering
\begin{tabular}{cccccc}
\hline
Method         & BCS$\uparrow$        & CSD$\downarrow$      & BHS$\uparrow$      & HSD$\downarrow$    & F1$\uparrow$     \\ \hline
LORIS w/o V    & 53.2                 & 28.2                 & 46.4               & 21.5               & 49.6             \\
LORIS w/o R    & 51.2                 & 29.8                 & 41.4               & 24.7               & 45.8             \\
LORIS w/ $V_m$ & 51.7                 & 23.5                 & \textbf{74.4}      & \textbf{19.0}      & 61.0             \\ \hline
LORIS (full)   & \textbf{58.8}        & \textbf{19.4}        & 67.1               & 21.1               & \textbf{62.7}    \\ \hline
\end{tabular}
\caption{Ablation studies on the engaged conditions on the LORIS$_{FE25}$ subset. M denotes rhythm conditioning, V indicates visual conditioning, and $V_m$ is a visual encoder that integrates RGB features and motion features. }
\vspace{-2mm}
\label{table:abcond}
\end{table}

\noindent\textbf{Engaged Conditioning.} We first investigate the role of data modalities that engage into conditional diffusion. We put forward three variants where `LORIS w/o V' and `LORIS w/o R' drop the visual and rhythm conditioning, respectively, and `LORIS w/ $V_m$' denotes using a trainable ST-GCN~\cite{yan2018spatial} network to encode human poses and then concatenate the extracted motion features with I3D features as the visual conditioning. Results are shown in Table~\ref{table:abcond}, which witnesses significant performance drops both during the absence of the visual and rhythm conditionings. This suggests that visual and rhythm conditionings are indispensable due to their grants for visual appearance and motion cues. Though LORIS w/ $V_m$ outperforms our full model in BHS and HSD, simply using RGB embeddings as visual encoder results in higher coverage scores and better overall performance (F1 scores). We argue that motion clues have already been taken into account during the extraction of visual rhythms, thus the motion embeddings are duplicated and can be omitted to lower the computation.

\begin{table}[htbp]
\small
\centering
\begin{tabular}{cccccc}
\hline
Method           & BCS                  & CSD                  & BHS                & HSD                & F1               \\ \hline
LORIS w/o LSTM   & 53.2                 & 25.3                 & \textbf{68.9}      & 21.2               & 60.0             \\
LORIS w/o Hawkes & 51.3                 & 26.2                 & 59.0               & 21.2               & 54.9             \\
LORIS w/o PE     & 52.4                 & 25.6                 & 64.0               & 21.6               & 57.6             \\
LORIS w/ rhymul  & 52.2                 & 25.2                 & 67.1               & 21.2               & 58.7             \\ \hline
LORIS (full)     & \textbf{58.8}        & \textbf{19.4}        & 67.1               & \textbf{21.1}      & \textbf{62.7}    \\ \hline
\end{tabular}
\caption{Ablation studies on model architecture designs on the LORIS$_{FE25}$ subset. PE denotes positional encoding of visual rhythm embedding and \textit{rhymul} is a variant that directly multiplies the rhythm envelope curves by the latent audio features.}
\label{table:abmod}
\end{table}

\noindent\textbf{Model Architecture.} We further evaluate the necessity of each model component, and results are shown in Table~\ref{table:abmod}. To investigate the effectiveness of temporal modeling, we first drop the Bi-LSTM layer termed `LORIS w/o LSTM'. Results show that though the ablated model achieves comparable performance on BHS, the coverage scores decline significantly (-5.6\%), indicating that temporal modeling for visual features is essential. We then analyze the impact of the rhythm conditioning module by ablating several variants: `LORIS w/o Hawkes' that removes the Hawkes process attached to the positional encoding, `LORIS w/o PE' that abolishing the entire positional encoding module, and 'LORIS w/ rhymul' that adopting a simpler conditioning strategy to directly multiply the visual rhythm envelopes by the auditory latent embedding (rather than using the cross-modality block). We observe that both positional encoding and the Hawkes process contribute to the rhythmic alignment. Besides, embed and cross-attend the rhythm peaks is a better conditioning approach than plain multiplication.

\subsection{Qualitative Results}

As illustrated in Figure~\ref{figure3}, we visualize music generated by LORIS together with the raw video-music pairs to illustrate rhythm correspondence and spectrogram analogousness. The upper part contrast visual rhythms with auditory rhythms extracted by ground-truth waveform and our generated music, where the synthesized contents reveal accurate alignment with visual appearances even when the ground-truth rhythms are mismatched (the 11th rhythm point). The lower part compares the log-melspectrograms of our generated music and ground truths. We find that the synthesized prosody patterns are alien to the raw audio since we do not add explicit reconstruction regularization to the raw audio sampling points, and the scholastic sampling nature of the diffusion sampling strategy guarantees the diversity of synthesized results. However, the position of audio peaks is likely to share a similar distribution, which indicates LORIS models the audio-visual correspondence and leverages such prior to generating music with comparable rhythm distribution. Besides, to show our model's ability in generating superior long-term soundtracks, we provide complementary qualitative demos in the \textbf{Supplementary Material}.

\section{Conclusion}
\label{Conclusion}
We have presented LORIS, a long-term rhythmic video soundtracker that generates video-conditioned musical waveforms via a context-aware latent diffusion model. A comprehensive benchmark for video soundtracks is also established, which includes a large-scale rhythmic video-music dataset varying from dancing to multiple sports events and a set of improved evaluation metrics. Experiments demonstrate that LORIS generates soundtracks with the best rhythm correspondence and satisfactory quality compared with existing methods. Nonetheless, LORIS only tackles fixed-length videos, thereby limiting its practicability, and the overall quality is still in need of improvement. In the future, we would seek different audio generation backbones for better musical quality and try context-aware modules to achieve unconstrained, even real-time generation purposes.

\section*{Acknowledgements}
This work is partially supported by the Shanghai Committee of Science and Technology (Grant No. 21DZ1100100). This work was supported in part by the National Natural Science Foundation of China under Grants 62102150.

\bibliography{reference}

\begin{thebibliography}{69}
\providecommand{\natexlab}[1]{#1}
\providecommand{\url}[1]{\texttt{#1}}
\expandafter\ifx\csname urlstyle\endcsname\relax
  \providecommand{\doi}[1]{doi: #1}\else
  \providecommand{\doi}{doi: \begingroup \urlstyle{rm}\Url}\fi

\bibitem[Agostinelli et~al.(2023)Agostinelli, Denk, Borsos, Engel, Verzetti,
  Caillon, Huang, Jansen, Roberts, Tagliasacchi,
  et~al.]{agostinelli2023musiclm}
Agostinelli, A., Denk, T.~I., Borsos, Z., Engel, J., Verzetti, M., Caillon, A.,
  Huang, Q., Jansen, A., Roberts, A., Tagliasacchi, M., et~al.
\newblock Musiclm: Generating music from text.
\newblock \emph{arXiv preprint arXiv:2301.11325}, 2023.

\bibitem[B{\"o}ck et~al.(2012)B{\"o}ck, Krebs, and Schedl]{bock2012evaluating}
B{\"o}ck, S., Krebs, F., and Schedl, M.
\newblock Evaluating the online capabilities of onset detection methods.
\newblock In \emph{ISMIR}, pp.\  49--54, 2012.

\bibitem[Brunner et~al.(2018)Brunner, Konrad, Wang, and
  Wattenhofer]{brunner2018midi}
Brunner, G., Konrad, A., Wang, Y., and Wattenhofer, R.
\newblock Midi-vae: Modeling dynamics and instrumentation of music with
  applications to style transfer.
\newblock \emph{arXiv preprint arXiv:1809.07600}, 2018.

\bibitem[Caillon \& Esling(2021)Caillon and Esling]{caillon2021rave}
Caillon, A. and Esling, P.
\newblock Rave: A variational autoencoder for fast and high-quality neural
  audio synthesis.
\newblock \emph{arXiv preprint arXiv:2111.05011}, 2021.

\bibitem[Carreira \& Zisserman(2017)Carreira and Zisserman]{carreira2017quo}
Carreira, J. and Zisserman, A.
\newblock Quo vadis, action recognition? a new model and the kinetics dataset.
\newblock In \emph{proceedings of the IEEE Conference on Computer Vision and
  Pattern Recognition}, pp.\  6299--6308, 2017.

\bibitem[Contributors(2020)]{mmpose2020}
Contributors, M.
\newblock Openmmlab pose estimation toolbox and benchmark.
\newblock \url{https://github.com/open-mmlab/mmpose}, 2020.

\bibitem[Davis \& Agrawala(2018)Davis and Agrawala]{davis2018visual}
Davis, A. and Agrawala, M.
\newblock Visual rhythm and beat.
\newblock In \emph{Proceedings of the IEEE Conference on Computer Vision and
  Pattern Recognition Workshops}, pp.\  2532--2535, 2018.

\bibitem[Dhariwal et~al.(2020)Dhariwal, Jun, Payne, Kim, Radford, and
  Sutskever]{dhariwal2020jukebox}
Dhariwal, P., Jun, H., Payne, C., Kim, J.~W., Radford, A., and Sutskever, I.
\newblock Jukebox: A generative model for music.
\newblock \emph{arXiv preprint arXiv:2005.00341}, 2020.

\bibitem[Di et~al.(2021)Di, Jiang, Liu, Wang, Zhu, He, Liu, and
  Yan]{di2021video}
Di, S., Jiang, Z., Liu, S., Wang, Z., Zhu, L., He, Z., Liu, H., and Yan, S.
\newblock Video background music generation with controllable music
  transformer.
\newblock In \emph{Proceedings of the 29th ACM International Conference on
  Multimedia}, pp.\  2037--2045, 2021.

\bibitem[Dong et~al.(2018)Dong, Hsiao, Yang, and Yang]{dong2018musegan}
Dong, H.-W., Hsiao, W.-Y., Yang, L.-C., and Yang, Y.-H.
\newblock Musegan: Multi-track sequential generative adversarial networks for
  symbolic music generation and accompaniment.
\newblock In \emph{Proceedings of the AAAI Conference on Artificial
  Intelligence}, volume~32, 2018.

\bibitem[Dong et~al.(2022)Dong, Chen, Dubnov, McAuley, and
  Berg-Kirkpatrick]{dong2022multitrack}
Dong, H.-W., Chen, K., Dubnov, S., McAuley, J., and Berg-Kirkpatrick, T.
\newblock Multitrack music transformer: Learning long-term dependencies in
  music with diverse instruments.
\newblock \emph{arXiv preprint arXiv:2207.06983}, 2022.

\bibitem[Gan et~al.(2020)Gan, Huang, Chen, Tenenbaum, and
  Torralba]{gan2020foley}
Gan, C., Huang, D., Chen, P., Tenenbaum, J.~B., and Torralba, A.
\newblock Foley music: Learning to generate music from videos.
\newblock In \emph{European Conference on Computer Vision}, pp.\  758--775.
  Springer, 2020.

\bibitem[Gu et~al.(2022)Gu, Chen, Bao, Wen, Zhang, Chen, Yuan, and
  Guo]{gu2022vector}
Gu, S., Chen, D., Bao, J., Wen, F., Zhang, B., Chen, D., Yuan, L., and Guo, B.
\newblock Vector quantized diffusion model for text-to-image synthesis.
\newblock In \emph{Proceedings of the IEEE/CVF Conference on Computer Vision
  and Pattern Recognition}, pp.\  10696--10706, 2022.

\bibitem[Hawkes(1971)]{hawkes1971spectra}
Hawkes, A.~G.
\newblock Spectra of some self-exciting and mutually exciting point processes.
\newblock \emph{Biometrika}, 58\penalty0 (1):\penalty0 83--90, 1971.

\bibitem[Hawthorne et~al.(2022)Hawthorne, Simon, Roberts, Zeghidour, Gardner,
  Manilow, and Engel]{hawthorne2022multi}
Hawthorne, C., Simon, I., Roberts, A., Zeghidour, N., Gardner, J., Manilow, E.,
  and Engel, J.
\newblock Multi-instrument music synthesis with spectrogram diffusion.
\newblock \emph{arXiv preprint arXiv:2206.05408}, 2022.

\bibitem[Hennequin et~al.(2020)Hennequin, Khlif, Voituret, and
  Moussallam]{spleeter2020}
Hennequin, R., Khlif, A., Voituret, F., and Moussallam, M.
\newblock Spleeter: a fast and efficient music source separation tool with
  pre-trained models.
\newblock \emph{Journal of Open Source Software}, 5\penalty0 (50):\penalty0
  2154, 2020.
\newblock \doi{10.21105/joss.02154}.
\newblock URL \url{https://doi.org/10.21105/joss.02154}.
\newblock Deezer Research.

\bibitem[Ho \& Salimans(2022)Ho and Salimans]{ho2022classifier}
Ho, J. and Salimans, T.
\newblock Classifier-free diffusion guidance.
\newblock \emph{arXiv preprint arXiv:2207.12598}, 2022.

\bibitem[Ho et~al.(2020)Ho, Jain, and Abbeel]{ho2020denoising}
Ho, J., Jain, A., and Abbeel, P.
\newblock Denoising diffusion probabilistic models.
\newblock \emph{Advances in Neural Information Processing Systems},
  33:\penalty0 6840--6851, 2020.

\bibitem[Hochreiter \& Schmidhuber(1997)Hochreiter and
  Schmidhuber]{hochreiter1997long}
Hochreiter, S. and Schmidhuber, J.
\newblock Long short-term memory.
\newblock \emph{Neural computation}, 9\penalty0 (8):\penalty0 1735--1780, 1997.

\bibitem[Hong et~al.(2022)Hong, Ding, Zheng, Liu, and Tang]{hong2022cogvideo}
Hong, W., Ding, M., Zheng, W., Liu, X., and Tang, J.
\newblock Cogvideo: Large-scale pretraining for text-to-video generation via
  transformers.
\newblock \emph{arXiv preprint arXiv:2205.15868}, 2022.

\bibitem[Huang et~al.(2018)Huang, Vaswani, Uszkoreit, Shazeer, Simon,
  Hawthorne, Dai, Hoffman, Dinculescu, and Eck]{huang2018music}
Huang, C.-Z.~A., Vaswani, A., Uszkoreit, J., Shazeer, N., Simon, I., Hawthorne,
  C., Dai, A.~M., Hoffman, M.~D., Dinculescu, M., and Eck, D.
\newblock Music transformer.
\newblock \emph{arXiv preprint arXiv:1809.04281}, 2018.

\bibitem[Huang et~al.(2023)Huang, Park, Wang, Denk, Ly, Chen, Zhang, Zhang, Yu,
  Frank, et~al.]{huang2023noise2music}
Huang, Q., Park, D.~S., Wang, T., Denk, T.~I., Ly, A., Chen, N., Zhang, Z.,
  Zhang, Z., Yu, J., Frank, C., et~al.
\newblock Noise2music: Text-conditioned music generation with diffusion models.
\newblock \emph{arXiv preprint arXiv:2302.03917}, 2023.

\bibitem[Huang \& Yang(2020)Huang and Yang]{huang2020pop}
Huang, Y.-S. and Yang, Y.-H.
\newblock Pop music transformer: Beat-based modeling and generation of
  expressive pop piano compositions.
\newblock In \emph{Proceedings of the 28th ACM International Conference on
  Multimedia}, pp.\  1180--1188, 2020.

\bibitem[Karras et~al.(2022)Karras, Aittala, Aila, and
  Laine]{karras2022elucidating}
Karras, T., Aittala, M., Aila, T., and Laine, S.
\newblock Elucidating the design space of diffusion-based generative models.
\newblock \emph{arXiv preprint arXiv:2206.00364}, 2022.

\bibitem[Kay et~al.(2017)Kay, Carreira, Simonyan, Zhang, Hillier,
  Vijayanarasimhan, Viola, Green, Back, Natsev, et~al.]{kay2017kinetics}
Kay, W., Carreira, J., Simonyan, K., Zhang, B., Hillier, C., Vijayanarasimhan,
  S., Viola, F., Green, T., Back, T., Natsev, P., et~al.
\newblock The kinetics human action video dataset.
\newblock \emph{arXiv preprint arXiv:1705.06950}, 2017.

\bibitem[Kreuk et~al.(2022)Kreuk, Synnaeve, Polyak, Singer, D{\'e}fossez,
  Copet, Parikh, Taigman, and Adi]{kreuk2022audiogen}
Kreuk, F., Synnaeve, G., Polyak, A., Singer, U., D{\'e}fossez, A., Copet, J.,
  Parikh, D., Taigman, Y., and Adi, Y.
\newblock Audiogen: Textually guided audio generation.
\newblock \emph{arXiv preprint arXiv:2209.15352}, 2022.

\bibitem[Kumar et~al.(2019)Kumar, Kumar, de~Boissiere, Gestin, Teoh, Sotelo,
  de~Br{\'e}bisson, Bengio, and Courville]{kumar2019melgan}
Kumar, K., Kumar, R., de~Boissiere, T., Gestin, L., Teoh, W.~Z., Sotelo, J.,
  de~Br{\'e}bisson, A., Bengio, Y., and Courville, A.~C.
\newblock Melgan: Generative adversarial networks for conditional waveform
  synthesis.
\newblock \emph{Advances in neural information processing systems}, 32, 2019.

\bibitem[Lee et~al.(2019)Lee, Yang, Liu, Wang, Lu, Yang, and
  Kautz]{lee2019dancing}
Lee, H.-Y., Yang, X., Liu, M.-Y., Wang, T.-C., Lu, Y.-D., Yang, M.-H., and
  Kautz, J.
\newblock Dancing to music.
\newblock \emph{Advances in neural information processing systems}, 32, 2019.

\bibitem[Li et~al.(2021)Li, Yang, Ross, and Kanazawa]{li2021ai}
Li, R., Yang, S., Ross, D.~A., and Kanazawa, A.
\newblock Ai choreographer: Music conditioned 3d dance generation with aist++.
\newblock In \emph{Proceedings of the IEEE/CVF International Conference on
  Computer Vision}, pp.\  13401--13412, 2021.

\bibitem[Lin et~al.(2014)Lin, Maire, Belongie, Hays, Perona, Ramanan,
  Doll{\'a}r, and Zitnick]{lin2014microsoft}
Lin, T.-Y., Maire, M., Belongie, S., Hays, J., Perona, P., Ramanan, D.,
  Doll{\'a}r, P., and Zitnick, C.~L.
\newblock Microsoft coco: Common objects in context.
\newblock In \emph{European conference on computer vision}, pp.\  740--755.
  Springer, 2014.

\bibitem[Loshchilov \& Hutter(2017)Loshchilov and
  Hutter]{loshchilov2017decoupled}
Loshchilov, I. and Hutter, F.
\newblock Decoupled weight decay regularization.
\newblock \emph{arXiv preprint arXiv:1711.05101}, 2017.

\bibitem[Mei \& Eisner(2017)Mei and Eisner]{mei2017neural}
Mei, H. and Eisner, J.~M.
\newblock The neural hawkes process: A neurally self-modulating multivariate
  point process.
\newblock \emph{Advances in neural information processing systems}, 30, 2017.

\bibitem[Mittal et~al.(2021)Mittal, Engel, Hawthorne, and
  Simon]{mittal2021symbolic}
Mittal, G., Engel, J., Hawthorne, C., and Simon, I.
\newblock Symbolic music generation with diffusion models.
\newblock \emph{arXiv preprint arXiv:2103.16091}, 2021.

\bibitem[Oord et~al.(2016)Oord, Dieleman, Zen, Simonyan, Vinyals, Graves,
  Kalchbrenner, Senior, and Kavukcuoglu]{oord2016wavenet}
Oord, A. v.~d., Dieleman, S., Zen, H., Simonyan, K., Vinyals, O., Graves, A.,
  Kalchbrenner, N., Senior, A., and Kavukcuoglu, K.
\newblock Wavenet: A generative model for raw audio.
\newblock \emph{arXiv preprint arXiv:1609.03499}, 2016.

\bibitem[Pasini \& Schl{\"u}ter(2022)Pasini and Schl{\"u}ter]{pasini2022musika}
Pasini, M. and Schl{\"u}ter, J.
\newblock Musika! fast infinite waveform music generation.
\newblock \emph{arXiv preprint arXiv:2208.08706}, 2022.

\bibitem[Pedersoli \& Goto(2020)Pedersoli and Goto]{pedersoli2020dance}
Pedersoli, F. and Goto, M.
\newblock Dance beat tracking from visual information alone.
\newblock In \emph{ISMIR}, pp.\  400--408, 2020.

\bibitem[Poole et~al.(2022)Poole, Jain, Barron, and
  Mildenhall]{poole2022dreamfusion}
Poole, B., Jain, A., Barron, J.~T., and Mildenhall, B.
\newblock Dreamfusion: Text-to-3d using 2d diffusion.
\newblock \emph{arXiv preprint arXiv:2209.14988}, 2022.

\bibitem[Radford et~al.(2021)Radford, Kim, Hallacy, Ramesh, Goh, Agarwal,
  Sastry, Askell, Mishkin, Clark, et~al.]{radford2021learning}
Radford, A., Kim, J.~W., Hallacy, C., Ramesh, A., Goh, G., Agarwal, S., Sastry,
  G., Askell, A., Mishkin, P., Clark, J., et~al.
\newblock Learning transferable visual models from natural language
  supervision.
\newblock In \emph{International Conference on Machine Learning}, pp.\
  8748--8763. PMLR, 2021.

\bibitem[Ramesh et~al.(2022)Ramesh, Dhariwal, Nichol, Chu, and
  Chen]{ramesh2022hierarchical}
Ramesh, A., Dhariwal, P., Nichol, A., Chu, C., and Chen, M.
\newblock Hierarchical text-conditional image generation with clip latents.
\newblock \emph{arXiv preprint arXiv:2204.06125}, 2022.

\bibitem[Ren et~al.(2020)Ren, He, Tan, Qin, Zhao, and Liu]{ren2020popmag}
Ren, Y., He, J., Tan, X., Qin, T., Zhao, Z., and Liu, T.-Y.
\newblock Popmag: Pop music accompaniment generation.
\newblock In \emph{Proceedings of the 28th ACM International Conference on
  Multimedia}, pp.\  1198--1206, 2020.

\bibitem[Roberts et~al.(2018)Roberts, Engel, Raffel, Hawthorne, and
  Eck]{roberts2018hierarchical}
Roberts, A., Engel, J., Raffel, C., Hawthorne, C., and Eck, D.
\newblock A hierarchical latent vector model for learning long-term structure
  in music.
\newblock In \emph{International conference on machine learning}, pp.\
  4364--4373. PMLR, 2018.

\bibitem[Rombach et~al.(2022)Rombach, Blattmann, Lorenz, Esser, and
  Ommer]{rombach2022high}
Rombach, R., Blattmann, A., Lorenz, D., Esser, P., and Ommer, B.
\newblock High-resolution image synthesis with latent diffusion models.
\newblock In \emph{Proceedings of the IEEE/CVF Conference on Computer Vision
  and Pattern Recognition}, pp.\  10684--10695, 2022.

\bibitem[Ronneberger et~al.(2015)Ronneberger, Fischer, and
  Brox]{ronneberger2015u}
Ronneberger, O., Fischer, P., and Brox, T.
\newblock U-net: Convolutional networks for biomedical image segmentation.
\newblock In \emph{International Conference on Medical image computing and
  computer-assisted intervention}, pp.\  234--241. Springer, 2015.

\bibitem[Schneider(2023)]{schneider2023archisound}
Schneider, F.
\newblock Archisound: Audio generation with diffusion.
\newblock \emph{arXiv preprint arXiv:2301.13267}, 2023.

\bibitem[Schneider et~al.(2023)Schneider, Jin, and
  Sch{\"o}lkopf]{schneider2023mo}
Schneider, F., Jin, Z., and Sch{\"o}lkopf, B.
\newblock Mo$\backslash$\^{} usai: Text-to-music generation with long-context
  latent diffusion.
\newblock \emph{arXiv preprint arXiv:2301.11757}, 2023.

\bibitem[Shao et~al.(2020)Shao, Zhao, Dai, and Lin]{shao2020finegym}
Shao, D., Zhao, Y., Dai, B., and Lin, D.
\newblock Finegym: A hierarchical video dataset for fine-grained action
  understanding.
\newblock In \emph{Proceedings of the IEEE/CVF conference on computer vision
  and pattern recognition}, pp.\  2616--2625, 2020.

\bibitem[Sheffer \& Adi(2022)Sheffer and Adi]{sheffer2022hear}
Sheffer, R. and Adi, Y.
\newblock I hear your true colors: Image guided audio generation.
\newblock \emph{arXiv preprint arXiv:2211.03089}, 2022.

\bibitem[Singer et~al.(2022)Singer, Polyak, Hayes, Yin, An, Zhang, Hu, Yang,
  Ashual, Gafni, et~al.]{singer2022make}
Singer, U., Polyak, A., Hayes, T., Yin, X., An, J., Zhang, S., Hu, Q., Yang,
  H., Ashual, O., Gafni, O., et~al.
\newblock Make-a-video: Text-to-video generation without text-video data.
\newblock \emph{arXiv preprint arXiv:2209.14792}, 2022.

\bibitem[Su et~al.(2020{\natexlab{a}})Su, Liu, and Shlizerman]{su2020audeo}
Su, K., Liu, X., and Shlizerman, E.
\newblock Audeo: Audio generation for a silent performance video.
\newblock \emph{Advances in Neural Information Processing Systems},
  33:\penalty0 3325--3337, 2020{\natexlab{a}}.

\bibitem[Su et~al.(2020{\natexlab{b}})Su, Liu, and Shlizerman]{su2020multi}
Su, K., Liu, X., and Shlizerman, E.
\newblock Multi-instrumentalist net: Unsupervised generation of music from body
  movements.
\newblock \emph{arXiv preprint arXiv:2012.03478}, 2020{\natexlab{b}}.

\bibitem[Su et~al.(2021)Su, Liu, and Shlizerman]{su2021does}
Su, K., Liu, X., and Shlizerman, E.
\newblock How does it sound? generation of rhythmic soundtracks for human
  movement videos.
\newblock In \emph{Conf. Neural Inf. Process. Syst}, volume~35, pp.\  0--10,
  2021.

\bibitem[Sun et~al.(2019)Sun, Xiao, Liu, and Wang]{sun2019deep}
Sun, K., Xiao, B., Liu, D., and Wang, J.
\newblock Deep high-resolution representation learning for human pose
  estimation.
\newblock In \emph{Proceedings of the IEEE/CVF conference on computer vision
  and pattern recognition}, pp.\  5693--5703, 2019.

\bibitem[Tang et~al.(2022)Tang, Gu, Bao, Chen, and Wen]{tang2022improved}
Tang, Z., Gu, S., Bao, J., Chen, D., and Wen, F.
\newblock Improved vector quantized diffusion models.
\newblock \emph{arXiv preprint arXiv:2205.16007}, 2022.

\bibitem[Tsuchida et~al.(2019)Tsuchida, Fukayama, Hamasaki, and
  Goto]{tsuchida2019aist}
Tsuchida, S., Fukayama, S., Hamasaki, M., and Goto, M.
\newblock Aist dance video database: Multi-genre, multi-dancer, and
  multi-camera database for dance information processing.
\newblock In \emph{ISMIR}, volume~1, pp.\ ~6, 2019.

\bibitem[Vaswani et~al.(2017)Vaswani, Shazeer, Parmar, Uszkoreit, Jones, Gomez,
  Kaiser, and Polosukhin]{vaswani2017attention}
Vaswani, A., Shazeer, N., Parmar, N., Uszkoreit, J., Jones, L., Gomez, A.~N.,
  Kaiser, {\L}., and Polosukhin, I.
\newblock Attention is all you need.
\newblock \emph{Advances in neural information processing systems}, 30, 2017.

\bibitem[von R{\"u}tte et~al.(2022)von R{\"u}tte, Biggio, Kilcher, and
  Hoffman]{von2022figaro}
von R{\"u}tte, D., Biggio, L., Kilcher, Y., and Hoffman, T.
\newblock Figaro: Generating symbolic music with fine-grained artistic control.
\newblock \emph{arXiv preprint arXiv:2201.10936}, 2022.

\bibitem[Wang et~al.(2020)Wang, Fang, and Zhao]{wang2020alignnet}
Wang, J., Fang, Z., and Zhao, H.
\newblock Alignnet: A unifying approach to audio-visual alignment.
\newblock In \emph{Proceedings of the IEEE/CVF Winter Conference on
  Applications of Computer Vision}, pp.\  3309--3317, 2020.

\bibitem[Xia et~al.(2022)Xia, Zhuge, Geng, Fan, Wei, He, and
  Zheng]{xia2022skating}
Xia, J., Zhuge, M., Geng, T., Fan, S., Wei, Y., He, Z., and Zheng, F.
\newblock Skating-mixer: Multimodal mlp for scoring figure skating.
\newblock \emph{arXiv preprint arXiv:2203.03990}, 2022.

\bibitem[Xie et~al.(2022)Xie, Zhang, Li, Tang, Du, and Hu]{xie2022vector}
Xie, P., Zhang, Q., Li, Z., Tang, H., Du, Y., and Hu, X.
\newblock Vector quantized diffusion model with codeunet for text-to-sign pose
  sequences generation.
\newblock \emph{arXiv preprint arXiv:2208.09141}, 2022.

\bibitem[Xie et~al.(2019)Xie, Wang, Hao, and Xu]{xie2019visual}
Xie, Y., Wang, H., Hao, Y., and Xu, Z.
\newblock Visual rhythm prediction with feature-aligning network.
\newblock In \emph{2019 16th International Conference on Machine Vision
  Applications (MVA)}, pp.\  1--6. IEEE, 2019.

\bibitem[Xu et~al.(2019)Xu, Fu, Zhang, Chen, Jiang, and Xue]{xu2019learning}
Xu, C., Fu, Y., Zhang, B., Chen, Z., Jiang, Y.-G., and Xue, X.
\newblock Learning to score figure skating sport videos.
\newblock \emph{IEEE transactions on circuits and systems for video
  technology}, 30\penalty0 (12):\penalty0 4578--4590, 2019.

\bibitem[Yan et~al.(2018)Yan, Xiong, and Lin]{yan2018spatial}
Yan, S., Xiong, Y., and Lin, D.
\newblock Spatial temporal graph convolutional networks for skeleton-based
  action recognition.
\newblock In \emph{Thirty-second AAAI conference on artificial intelligence},
  2018.

\bibitem[Yang et~al.(2022)Yang, Yu, Wang, Wang, Weng, Zou, and
  Yu]{yang2022diffsound}
Yang, D., Yu, J., Wang, H., Wang, W., Weng, C., Zou, Y., and Yu, D.
\newblock Diffsound: Discrete diffusion model for text-to-sound generation.
\newblock \emph{arXiv preprint arXiv:2207.09983}, 2022.

\bibitem[Yu et~al.(2022)Yu, Pu, Cheng, Feng, and Shan]{yu2022self}
Yu, J., Pu, J., Cheng, Y., Feng, R., and Shan, Y.
\newblock Self-supervised learning of music-dance representation through
  explicit-implicit rhythm synchronization.
\newblock \emph{arXiv preprint arXiv:2207.03190}, 2022.

\bibitem[Zhang et~al.(2020{\natexlab{a}})Zhang, Shen, Xu, and
  Shen]{zhang2020temporal}
Zhang, J., Shen, F., Xu, X., and Shen, H.~T.
\newblock Temporal reasoning graph for activity recognition.
\newblock \emph{IEEE Transactions on Image Processing}, 29:\penalty0
  5491--5506, 2020{\natexlab{a}}.

\bibitem[Zhang et~al.(2020{\natexlab{b}})Zhang, Lipani, Kirnap, and
  Yilmaz]{zhang2020self}
Zhang, Q., Lipani, A., Kirnap, O., and Yilmaz, E.
\newblock Self-attentive hawkes process.
\newblock In \emph{International conference on machine learning}, pp.\
  11183--11193. PMLR, 2020{\natexlab{b}}.

\bibitem[Zhang et~al.(2022)Zhang, Zhang, Shao, Shan, and Xia]{zhang2022vis2mus}
Zhang, R., Zhang, Y., Shao, K., Shan, Y., and Xia, G.
\newblock Vis2mus: Exploring multimodal representation mapping for controllable
  music generation.
\newblock \emph{arXiv preprint arXiv:2211.05543}, 2022.

\bibitem[Zhu et~al.(2022{\natexlab{a}})Zhu, Olszewski, Wu, Achlioptas, Chai,
  Yan, and Tulyakov]{zhu2022quantized}
Zhu, Y., Olszewski, K., Wu, Y., Achlioptas, P., Chai, M., Yan, Y., and
  Tulyakov, S.
\newblock Quantized gan for complex music generation from dance videos.
\newblock \emph{arXiv preprint arXiv:2204.00604}, 2022{\natexlab{a}}.

\bibitem[Zhu et~al.(2022{\natexlab{b}})Zhu, Wu, Olszewski, Ren, Tulyakov, and
  Yan]{zhu2022discrete}
Zhu, Y., Wu, Y., Olszewski, K., Ren, J., Tulyakov, S., and Yan, Y.
\newblock Discrete contrastive diffusion for cross-modal and conditional
  generation.
\newblock \emph{arXiv preprint arXiv:2206.07771}, 2022{\natexlab{b}}.

\end{thebibliography}
\bibliographystyle{icml2023}

\clearpage
\appendix
\section{Model Details}

\begin{figure*}[ht]
\centering
\includegraphics[width=0.98\textwidth]{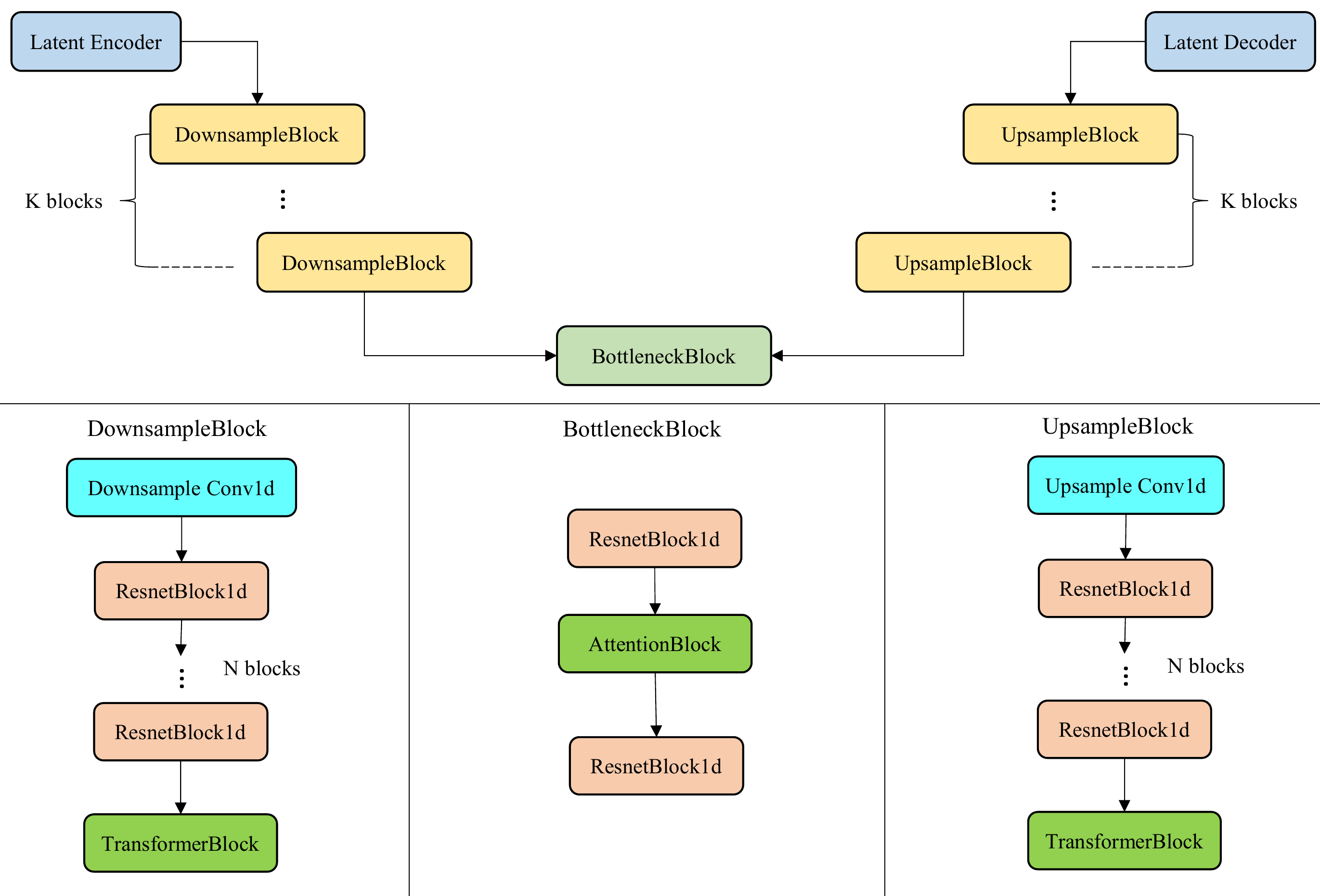}
\caption{Illustration of the audio diffusion backbone that includes a latent encoder, u-net, and a latent decoder. The u-net consists of one bottleneck block and K stacking downsample and upsample blocks. The basic unit module name ResnetBlock1d involves several 1D convolutional layers, which form the main architecture of the u-net component.}
\label{appenfigure1}
\end{figure*}

We here give a brief introduction of the model architecture, including the illustration of the backbone structure and detailed descriptions of the scaling parameters of the denoiser.
\subsection{Diffusion Model Architecture}
Overview of the audio diffusion backbone~\cite{schneider2023archisound} is illustrated in the upper part of Figure~\ref{appenfigure1}, which mainly comprises the latent encoder, u-net~\cite{ronneberger2015u}, and latent decoder. The detailed structure of each module is illustrated in the lower part, where all components involve a basic unit named ResnetBlock1d. To be specific, ResnetBlock1d includes three 1D convolutional blocks that perform feature integration. Furthermore, all conditional embeddings are introduced into the diffusion process via the cross-modal attention block~\cite{vaswani2017attention} in all ResnetBlock1d modules of downsample block, upsample block, and bottleneck block, while the ResnetBlock1d modules in latent encoder and decoder do not involve any cross-modality interaction.

\subsection{Scaling Parameters}
\label{scalingparam}
We adopt EDM~\cite{karras2022elucidating} to ameliorate the denoiser $D_\theta(z, \sigma_t)$ via skip scaling $c_{skip}(\sigma_t)$, output scaling $c_{out}(\sigma_t)$, input scaling $c_{in}(\sigma_t)$, and noise mapping $c_{noise}(\sigma_t)$. The entire process can be formulated as:
\begin{equation}
D_{\theta}(z, \sigma_t) = c_{skip}(\sigma_t)z + c_{out}(\sigma_t)f_{\theta}(c_{in}(\sigma_t)z, c_{noise}(\sigma_t)),
\end{equation}
\begin{equation}
    c_{skip}(\sigma_t) = \sigma_{data}^2/(\sigma_t^2+\sigma_{data}^2)
\end{equation}
\begin{equation}
    c_{out}(\sigma_t) = \sigma_t \cdot \sigma_{data}/\sqrt{\sigma^2_{data}+\sigma_t^2}
\end{equation}
\begin{equation}
    c_{in}(\sigma_t)=1/\sqrt{\sigma_t^2+\sigma_{data}^2}
\end{equation}
\begin{equation}
    c_{noise}(\sigma_t)=\frac{1}{4}ln(\sigma_t).
\end{equation}
In practice, the parameter $\sigma_{data}$ is set to 0.1, and the noise distribution is set as $ln(\sigma) \sim \mathcal{N}(P_{mean}, P_{std}^2)$, where the mean and standard deviation parameters are set as $P_{mean}=-3.0, P_{std}=1.0$ .

\section{Additional Details}
In this section, we elaborate on additional experimental settings, dataset information, and details of the evaluation metrics.

\subsection{Experimental Settings}
We here complement additional implementation details. During training, we set the batch size of dancing and floor exercise videos as 10, and each figure skating batch involves 8 samples. For the diffusion probabilistic model backbone, we set the patch block number as 1 and the patch factor as 32. The multipliers of ResnetBlock1d for each upsample block and downsample block are (1, 2, 4, 4, 4, 4, 4), the factors and num\_blocks hyperparameter vectors for UNet1d module are (4, 4, 4, 2, 2, 2) and (2, 2, 2, 2, 2, 2), respectively. The head number for all attention blocks is set as 16, and we employ a 0.1 batch dropout probability for the conditional generation. We transform the monophonic input to binaural via duplicating the waveform data on the channel dimension to meet the prerequisite of the pre-trained audio-diffusion-pytorch~\cite{schneider2023archisound} model. Besides, we apply data augmentation during training by adding a uniformly-sampled acoustic amplitude to the raw audio sample points. We also normalize the input data and multiply it by a factor $f=0.95$ to stabilize training. It is also noted that different from previous methods~\cite{zhu2022discrete, zhu2022quantized}, we do not adopt any noise-reducing approach for soundtrack calibration.

\subsection{Dataset Information}
Since all video-music pairs are curated from existing datasets~\cite{zhu2022quantized, li2021ai, shao2020finegym, xia2022skating, xu2019learning}, we manually filter videos based on certain policies to guarantee analogous data distributions while maintaining adequate diversity. Concretely, we keep the character number as 1 and remove all videos with multiple persons, including group dances, pairs skating, and group gymnastics. Furthermore, all kinds of human voices (audience cheers, commentaries, song vocals, etc.) are removed by the sound source separation model Spleeter~\cite{spleeter2020}. For diversity, we try to involve more genders, races, and music types in all rhythmic scenarios. Furthermore, camera views in all video categories are diverse, where shot changes and scales exist in sports videos, and dances are captured via cameras from different angles.

\subsection{Evaluation Metrics}
\label{evamet}
As we mentioned in the main manuscript, parameters of audio onset detection algorithms~\cite{bock2012evaluating} are adaptively adjusted for the long-term scenario. To be specific, we modify the hyperparameters of the peak-picking strategy, where the previous and post distances of computing the local average are adjusted from 1, 1 to 2, 2, the previous and post distances of calculating local maximum are modified from 0, 1 to 3, 3, and the threshold $\delta$ of picking peaks is set to 0.2. We expand the window size of picking the local maximum, which is likely to have a higher value than the maximum in a smaller area. By this means, the selected local maximum has more opportunities to meet the requirements of the threshold $\delta$, thus leading to a more compact vector that includes more visual rhythm points.

The Mean Opinion Score (MOS) is a subjective metric obtained by computing the average scores of a user study regarding the overall musical quality. Concretely, 14 participants are required to listen to 25 musical waveforms generated by LORIS and four baseline methods (5 samples for each method), then give scores ranging from 1 to 5 to evaluate the overall musical quality. During the evaluation, participants are only provided with auditory samples while visual counterparts are unavailable. Hence, this subjective metric only measures the uni-modal musical quality, and the cross-modality rhythmic correspondence is evaluated by other objective metrics.

\section{Visual Rhythm Extractor}
\label{rhythmrationale}
In this section, we analyze the discrepancy between previous visual rhythm extraction approaches and interpret the rationale of our improved rule-based method. In total, all existing methods, including our improved version, share a common paradigm that captures the magnitude of human cadent movements based on various features. To be specific, \cite{davis2018visual} propose to utilize audio spectrogram and onset detection algorithms on optical flow features to extract visual rhythms, which first show the feasibility of monitoring the amplitude variety of visual cues. But two drawbacks emerge: 1. The extraction of optical flow is computationally-expensive when adopting a large sampling rate of video frames, which sets up an obstacle for the rhythm detection of high-quality videos. 2. The transformation among visual onset, tempograms, and visual beats is complicated, and leveraging the periodic visual beats for inter-modality rhythm alignment severely restrains flexibility. Therefore, it is more succinct to directly regard visual onset as visual rhythms to perform the subsequent cross-modal interaction. More recently, \cite{su2021does} adopt human poses rather than optical flows to conduct analogous motion saliency detection. Directograms are first computed based on the angular positions and magnitudes of human poses, and Short-Time-Fourier Transform (STFT) is performed on the first-order difference of directograms (kinematic offsets) to locate the visual rhythm points. Two major differences lie between this approach and our rule-based method. One is that we leverage the peak-picking strategy similar (yet different hyperparameters due to the modality discrepancy) to audio onset peak detection algorithm~\cite{bock2012evaluating} instead of STFT to remain similar peak-picking principles. The other is that we prefer the human motions than human poses as input for directogram calculation, since angles of poses $P(t, j, x, y)$ with the x-axis only reflect its relative position with respect to the reference coordinate system, while the angular position of human motion $M(t, j, x, y)$ measures the changing direction and velocity of human poses. Besides, some recent works leverage deep-learning methods to implicitly model visual rhythms~\cite{xie2019visual, wang2020alignnet, pedersoli2020dance}, or combine traditional algorithms with trainable networks~\cite{yu2022self}. Both of them are not employed in our framework in case that much more parameters are additionally introduced.
\begin{figure*}[ht]
\centering
\includegraphics[width=0.98\textwidth]{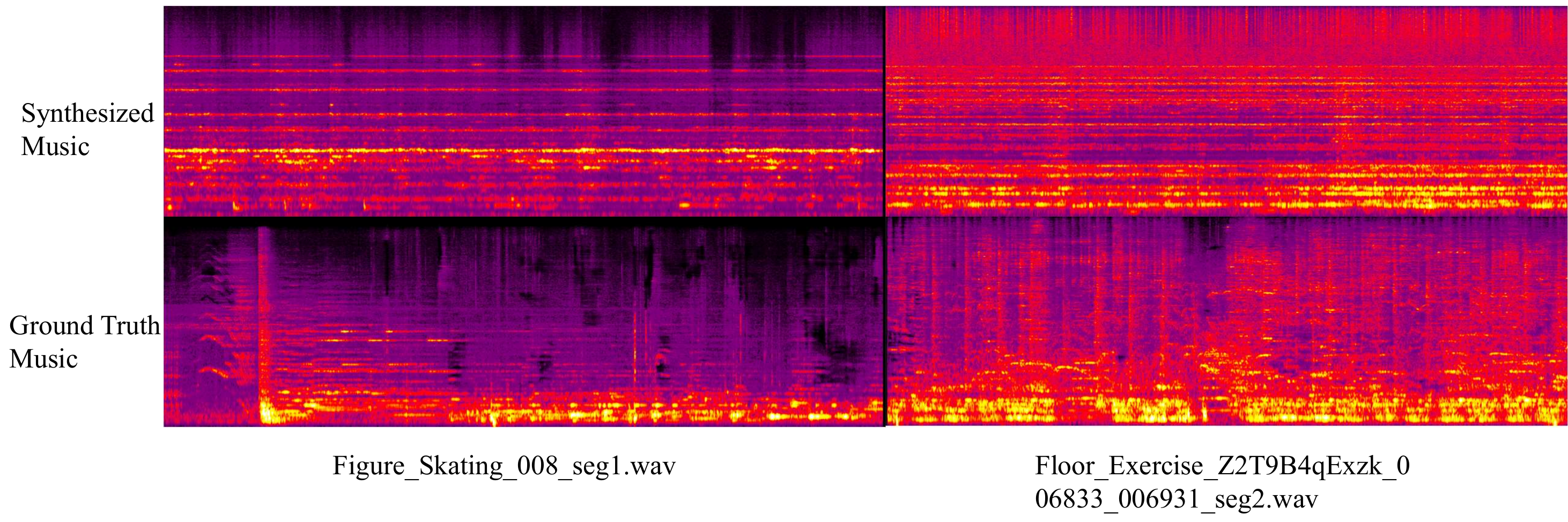}
\caption{Log-melspectrograms of two failure examples. The upper part denotes the synthesized music and the lower part indicates the ground-truth soundtracks.}
\label{appenfigure2}
\end{figure*}

\section{Additional Ablation Studies}
To fully probe the experimental settings of our proposed model, we conduct additional ablation studies on the sampling hyperparameters.  

\noindent\textbf{Sampling step.} We first investigate the impact of different sampling steps $s$ during inference. Results are shown in Table~\ref{table:abstep}, where the increase of sampling step number from 10 to 50 is accompanied by performance boosts. Furthermore, the performance of $s=50$ and $s=100$ is comparable, indicating that a much larger sampling step number cannot benefit generation quality. We hence set the sampling step number as 50 for computation simplicity.  
\begin{table}[htbp]
\small
\centering
\resizebox{\linewidth}{!}{
\begin{tabular}{lccccc}
\toprule
Step Num        & BCS              & CSD              & BHS            & HSD            & F1               \\ \midrule
s=10            & 52.5             & 26.4             & 59.5           & \textbf{17.9}  & 55.8             \\
s=20            & 51.1             & 25.6             & 61.2           & 20.6           & 55.7             \\
s=30            & 51.4             & 26.3             & 65.0           & 20.2           & 57.4             \\
s=100           & 58.3             & 20.0             & \textbf{68.2}  & 20.5           & \textbf{62.9}    \\ \midrule
s=50 (Ours)     & \textbf{58.8}    & \textbf{19.4}    & 67.1           & 21.1           & 62.7             \\ \bottomrule
\end{tabular}}
\caption{Ablation studies on the number of sampling steps $s$ on the LORIS$_{FE25}$ subset.}
\label{table:abstep}
\end{table}

\noindent\textbf{Guidance scale.} We further explore the influence of different guidance scales of the classifier-free diffusion guidance~\cite{ho2022classifier}. As shown in Table~\ref{table:abgui}, we conduct ablation studies on a large range of guidance scale options, and results show that setting $s=20$ brings optimal rhythmic correspondence performance. Such conclusions are also coherent with other cross-modality conditional scenarios such as text-to-image, where setting a larger guidance scale $w$ than uni-modal conditional generation witnesses a better generation quality.

\begin{table}[htbp]
\small
\centering
\resizebox{\linewidth}{!}{
\begin{tabular}{lccccc}
\toprule
Guidance Scale       & BCS              & CSD              & BHS            & HSD            & F1               \\ \midrule
w=2                  & 51.8             & 26.7             & 62.4           & 25.1           & 56.6             \\
w=5                  & 53.4             & 27.4             & 66.8           & \textbf{19.9}  & 59.4             \\
w=10                 & 53.2             & 23.9             & 61.8           & 24.5           & 57.2             \\
w=30                 & 57.9             & 22.1             & \textbf{68.1}  & 42.0           & 62.6             \\
w=50                 & 58.3             & 20.8             & 64.5           & 24.5           & 61.2             \\
w=100                & 57.9             & 23.9             & 62.7           & 27.2           & 60.2             \\ \midrule
w=20 (Ours)          & \textbf{58.8}    & \textbf{19.4}    & 67.1           & 21.1           & \textbf{62.7}    \\ \bottomrule
\end{tabular}}
\caption{Ablation studies on the guidance scale $w$ of classifier-free conditional generation on the LORIS$_{FE25}$ subset.}
\label{table:abgui}
\end{table}
\section{Failure Cases}

As depicted in Figure~\ref{appenfigure2}, we show the log-melspectrograms of two failure examples as well as their ground-truth counterparts. The left part is the soundtrack of a figure skating video, where the skater moves smoothly in sync with the gentle music. Though the musical quality is satisfactory, the changing amplitude of human motions is not significant, thus our model fails to correctly pick visual rhythmic peaks. This instance shows our model's incapability in synthesizing rhythm correlated soundtrack under certain circumstances. We argue that additional samples under such scenarios are required to perform targeted training. The right part denotes the soundtrack of a floor exercise video, where our model generates a musical waveform with extremely low quality. This suggests that the stability and robustness of our model are still in need of progress despite the fact that having equipped with a pre-trained audio diffusion backbone. Demos of all failure cases are provided in the \textbf{Supplementary Material.}

\section{Further Discussion}
\subsection{Limitation}

LORIS has two main limitations. One is that LORIS only tackles trimmed videos with fixed lengths, which hinders the proposed model to tackle unconstrained data or real-time soundtrack scenarios. One possible solution is to attach a sliding window on the untrimmed video as a segment-wise generation. However, certain modifications are required to construct the global correlation across segments to make the entire auditory sequence continuous and harmonious. The other limitation lies in the dependence on the pre-trained encoder. Analogous with previous methods~\cite{zhu2022discrete, zhu2022quantized} that utilize a vector-quantized variational autoencoder~\cite{dhariwal2020jukebox}, our method relies on the pre-trained latent diffusion model~\cite{schneider2023archisound}, resulting in the inflexibility of appending or replacing several particular components.

\subsection{Potential Social Impact}
Our generative model may be employed to produce fake videos with forgery soundtracks. Furthermore, equipped with DeepFake videos synthesized by video generative algorithms, a complete forgery audio-visual pair can be automatically fabricated, which may have potential negative impacts on the web environment. Hence, appropriate supervision is essential to guarantee a controllable and harmless generation.

\end{document}